\documentclass[12pt]{article}
\usepackage{latexsym,epsfig,graphicx,amsmath,amssymb,amscd,undertilde,multirow,chicago,psfrag,slashbox, paralist}
\newcommand{\thetavec}{{\boldsymbol{\theta}}}

\newcommand{\degreeC}{\ensuremath{^\circ}\text{C}}

\newcommand{\nm}{{\rm nm}}

\newcommand{\thetavechat}{\widehat{\thetavec}}

\newcommand{\Degpath}{\mathcal{G}}
\newcommand{\NOR}{{\rm N}}

\newcommand{\Lamp}{{\rm Lamp}}
\newcommand{\cm}{{\rm cm}}
\newcommand{\Temp}{{\rm Temp}}
\newcommand{\RH}{{\rm RH}}
\newcommand{\ND}{{\rm ND}}
\newcommand{\Filter}{{\rm Filter}}
\newcommand{\BP}{{\rm BP}}
\newcommand{\rh}{{\rm rh}}

\usepackage{hyphenat}

\begin{document}
\title{Development of an Accelerated Test Methodology to the Predict
  Service Life of Polymeric Materials Subject to Outdoor Weathering}

\author{Yuanyuan Duan$^1$, Yili Hong$^1$, William Q. Meeker$^2$,\\
Deborah L. Stanley$^3$, and Xiaohong Gu$^3$\\[1ex]
{\small $^1$Department of Statistics, Virginia Tech, Blacksburg, VA 24061}\\
{\small $^2$Department of Statistics, Iowa State University, Ames, IA 50011}\\
{\small $^3$Engineering Laboratory, National Institute of Standards and Technology,}\\ {\small Gaithersburg, MD 20899}
}
\date{\today}
\maketitle

\begin{abstract}
Service life prediction is of great importance to manufacturers of
coatings and other polymeric materials. Photodegradation, driven
primarily by ultraviolet (UV) radiation, is the primary cause of
failure for organic paints and coatings, as well as many other
products made from polymeric materials exposed to
sunlight. Traditional methods of service life prediction involve the
use of outdoor exposure in harsh UV environments (e.g., Florida and
Arizona). Such tests, however, require too much time (generally many years) to
do an evaluation. Non-scientific attempts to simply ``speed up the clock''
result in incorrect predictions. This paper describes the
statistical methods that were developed for a scientifically-based
approach to using laboratory accelerated tests to produce timely
predictions of outdoor service life. The approach involves careful
experimentation and identifying a physics/chemistry-motivated model
that will adequately describe photodegradation paths of polymeric
materials. The model incorporates the effects of explanatory
variables UV spectrum, UV intensity, temperature, and humidity. We
use a nonlinear mixed-effects model to describe the sample paths.
The methods are illustrated with accelerated laboratory test data
for a model epoxy coating. The validity of the methodology is checked
by extending our model to allow for dynamic covariates and
comparing predictions with specimens that were exposed in an outdoor
environment where the explanatory variables are uncontrolled but
recorded.


\end{abstract}

\newpage
\section{Introduction}
\subsection{Background and Motivation}
Polymeric materials are widely used in many products such as
paints, coatings, and components in systems such as photovoltaic
power generation equipment (e.g., encapsulant and
backsheet). Photodegradation caused by ultraviolet (UV) radiation is
the primary cause of failure for paints and coatings, as well as
many other products made from polymeric materials that are exposed
to sunlight. Other environmental variables including temperature and
humidity can also affect degradation rates. When a new product that
will be subjected to outdoor weathering is developed, it is
necessary to assess the product's service life. As an example, for
paints and coatings, the traditional method of service life
prediction involves sending perhaps ten coated panels to Florida
(where it is sunny and humid) and another ten panels to Arizona (where
it is sunny and dry). Then every six months one panel is returned
from each exposure location for detailed evaluation (e.g., to
quantify chemical and physical changes over time). If the amount of
degradation is sufficiently small after, say, five years, the
service life is deemed to satisfactorily long.

The problem with the traditional method of service life prediction
is that it takes too long to obtain the needed assessment. For many
decades, accelerated tests (e.g., \citeNP{Nelson1990}) have been
used successfully to assess the lifetime of products and components in
environments that do not involve UV exposure. Accelerated tests for
photodegradation are, however, more complicated. Non-scientific
approaches to achieve acceleration of the degradation process by
simply ``speeding up the clock'' in laboratory testing led to incorrect
predictions. It is believed that the efforts failed for a
combination of reasons including that UV lamps do not have the same
spectral irradiance distribution as the sun and that varying all
experimental factors simultaneously (the opposite of what would be
done in a carefully designed experiment) does not provide useful
information for modeling and prediction.

Scientists at the U.S. National Institute of Standards and
Technology (NIST), in collaboration with scientists and engineers
from companies and other organizations, conducted a multi-year
research program to develop a scientifically-based laboratory
accelerated testing methodology that could be used to predict the
service life of polymeric materials subjected to outdoor
weathering. The purpose of this paper is to describe the statistical
methods that were used for physical/chemical modeling and to compute
predictions of outdoor service life, based on the laboratory
accelerated test. The methods were validated by comparing the
predictions with specimens that were subjected to outdoor exposure
where dynamic explanatory variables (i.e., time-varying covariates)
although not controlled, were recorded.

The laboratory accelerated weathering tests were conducted using
the NIST SPHERE (Simulated Photodegradation via High Energy Radiant
Exposure), a device in which spectral UV wavelength, UV spectral
intensity, temperature, and relative humidity (RH) can be controlled
over time. Also, outdoor-exposure experiments were conducted on the roof
of a NIST building in Maryland over different time periods. Both
sets of experiments used a model epoxy coating. Chemical degradation
was measured on both the laboratory accelerated test specimens and
the outdoor-exposed specimens every few days using Fourier transform
infrared (FTIR) spectroscopy. Longitudinal information on ambient
temperature, RH, and the solar intensity and spectrum for outdoor-exposed
specimens were carefully recorded at 12-minute intervals over the
period of outdoor exposure.

\subsection{A General Framework for Degradation Prediction}
This section summarizes the major steps in the general framework for
accelerated  photodegradation testing of polymeric materials
that are subject to outdoor weathering.

\begin{inparaenum}
\item \emph{Use the accelerated test data and knowledge of the
  physics and chemistry of the degradation process to help identify
  the functional forms for the experimental variables as they relate to
  the degradation path model.}

\item \emph{Use the identified functional forms and the accelerated
  test data to build a degradation path model linking the sample
  degradation paths and the experimental variables.}

\item \emph{Use the identified model to generate predictions of
  degradation for a given covariate histories.}

\item \emph{To verify the effectiveness of the accelerated test
  methodology, compare predictions, based on the accelerated test
  degradation data and model, with observed degradation paths for
  outdoor-exposed specimens.}

\item \emph{Use prediction intervals to quantify the statistical
  uncertainties associated with the outdoor degradation predictions.}

\end{inparaenum}

In summary, the presented modeling approach advocates the strategy
of combining physical/chemical knowledge and accelerated test data
to build a model that can predict field performance.

\subsection{Related Literature and Contribution of This Work}
Traditional life tests generally require a long time to obtain a
substantial number of failures because modern products are designed
to last a long time. To overcome the time constraints of life tests,
degradation data provide quantitative measurements and thus more
information than failure data. \citeN{LuMeeker1993}, and
\citeN{HongMeekerEscobar2010} give examples of models and analyses
of degradation data. To speed up the degradation process and provide
information in a more timely manner, accelerated degradation tests
are commonly used (e.g., Chapter 12 of \citeNP{Nelson1990}, Chapter
21 of \citeNP{MeekerEscobar1998}, and
\citeNP{MeekerEscobarLu1998}). The use of degradation data in
accelerated tests provides more credible and precise reliability
estimates and a firmer basis for extrapolation at normal use
conditions. Potential accelerating variables include the use rate or
aging rate of a product, exposure intensity, voltage stress,
temperature, humidity, etc. (e.g.,
\citeNP{EscobarMeeker2006}). Combinations of these accelerating
variables are sometimes used.

For products and systems in the field, the degradation process
usually depends on dynamic environmental covariates. Dynamic data
collection is becoming much easier with modern sensor technology,
motivating the modeling of the effect of dynamic covariates. For
example, \citeN{HahnDoganaksoy2008} described sensors recording
dynamic covariates such as oil pressure and oil/water temperature in
locomotive engines and how the information could be used to diagnose
system faults. \shortciteN{Spurgeonetal2005} described an automatic
system that can monitor dissolved gas in the insulating oil in
high-voltage power transformers to detect the occurrence of arcing
that could, if not corrected, lead to a catastrophic
failure.

Degradation processes are often affected by dynamic
covariates but it is challenging to incorporate such information
into a degradation-process model. The cumulative damage model has
been used to describe the effect that dynamic covariates have on
degradation and failure-time processes (e.g., \citeNP{Nelson1990},
\citeNP{Subramanian1995}, \citeNP{BagdonaviciusNikulin2001},
\citeNP{VacaTrigoMeeker2009}, \citeNP{HongMeeker2010}, and
\citeNP{HongMeeker2010b}). \shortciteN{HongDuanMeekerGuStanley2015}, and \shortciteN{XuHongJin2016} use the NIST
outdoor-exposure data to build predictive models for
degradation. One problem with those approaches is that there is no
acceleration and the evaluation of service life, for many products,
would take too long.

It takes a long time to observe the actual service life performance
of new products and systems. For example, when new coatings are
being developed, there is a need to have extensive outdoor exposure to
characterize service life performance. Depending on the product,
such outdoor exposures could take many years or even decades
(\shortciteNP{Martin1996}). Also, the outdoor exposure conditions
are complicated, due to the joint effects of multiple
dynamic covariates. Laboratory accelerated tests can be conducted over much
shorter  periods of time. With knowledge of the failure mechanisms and
proper scientific modeling of data from a well-designed experiment,
it is possible to predict outdoor service life performance. It will
be practically useful if the laboratory accelerated test data and
the outdoor performance data can be linked. \shortciteN{Guetal2008}
described three potential approaches to link laboratory accelerated
degradation test data with outdoor-exposure data for a coating
system.

\begin{inparaitem}
\item
An approach based on chemical concentration ratios,

\item
A heuristic approach, and

\item
An approach based on a predictive model.
\end{inparaitem}

In a preliminary report of the NIST experimental program,
\citeN{VacaTrigoMeeker2009} described a predictive model to link the
NIST laboratory accelerated test data and outdoor-exposure
data. They used a nonlinear model for the accelerated test data and a
cumulative damage model to predict the outdoor-exposure data. In
this paper we extend this previous work to provide a better, more
scientifically justified model for the explanatory-variable effects
and the different sources of variability. We also provide improved
methods to make service life predictions and to quantify prediction
error, making it possible to quantitatively compare the predictions
with the outdoor-exposure data.

More specifically, in this paper we extend the work in
\citeN{VacaTrigoMeeker2009} by using a sophisticated nonlinear
mixed-effects model with careful physically-motivated modeling of
the effects of the accelerating variables on the sample degradation
paths. This improved model provides enhanced prediction performance
and the ability to quantify prediction uncertainty with prediction
intervals. More generally, this paper provides the following
advances.

\begin{inparaitem}
\item We propose and develop a general modeling and prediction
  framework for accelerated photodegradation testing of
  polymeric materials. The methodology framework is illustrated by
  the data collected from a model epoxy (which would tend to degrade
  rapidly---providing further acceleration) but it can also be used to
  predict the degradation of other materials such as ethylene-vinyl
  acetate (EVA) and polyethylene terephthalate (PET) and materials
  containing UV protection (similar to sunscreens that humans use to
  protect their skin from harmful UV radiation).

\item The proposed stage-wise modeling method provides a means to
  combine scientific knowledge of the degradation process with
  experimental data to identify the functional form of each
  accelerating variable in the degradation path model. This general
  methodology will be useful for other kinds of accelerated
  degradation studies.

\item A prediction procedure based on a cumulative damage model is
  developed and prediction uncertainties  are
  quantified with prediction intervals.

\end{inparaitem}

\subsection{Overview}
The rest of this paper is organized as
follows. Section~\ref{sec:pdata} describes the laboratory
accelerated test and the outdoor-exposure experiments and provides notation
for the data. Section~\ref{sec:photo.model} describes the nonlinear
mixed-effects model and defines total effective
dosage. Section~\ref{sec:modeling.indoor.data} uses the laboratory
accelerated test data to compute estimates of a categorical-effects
model, providing information about the functional forms of the
experimental variables needed to
identify a model relating photodegradation to the experimental
variables. Section~\ref{sec:pred.outdoor} uses model parameter
estimates from the laboratory accelerated test data and a cumulative
damage model to predict outdoor-exposure degradation and compares the
predictions with actual outdoor-exposure degradation paths. A comparison is
also done for several different models in terms of model fitting and
prediction accuracy. Section~\ref{sec:conclusion2} contains
conclusions and discussion of areas for future research.

\section{Photodegradation Time Scale and Data}\label{sec:pdata}
\subsection{Choice of a Time Scale}
As described on page 4 of \citeN{CoxOakes1984} and page 18 of
\citeN{MeekerEscobar1998}, when conducting any kind of failure-time
study, it is important to carefully consider the time scale to be
used. For example, roller bearing life would most reasonably be
measured in terms of something proportional to the number of
revolutions. If the bearing is installed in an automobile, that
information might not be available and so the number of miles driven
would be a useful surrogate. For a rubber seal or an adhesive in a controlled
environment and no UV exposure, something proportional to real time
(e.g., months in service) would be appropriate. For a coating
subjected to UV exposure, the scientifically appropriate time scale
would be proportional to the number of photons that get absorbed
into the coating, taking into account that shorter wavelength
photons are more energetic (and thus have a higher probability to
cause damage). For those who study photodegradation, such a measure
is called UV dosage, as will be described in detail in subsequent
sections of this paper.

\subsection{Laboratory Accelerated Test Experiments and Data}
\begin{figure}
\begin{center}
\includegraphics[width=.5\textwidth]{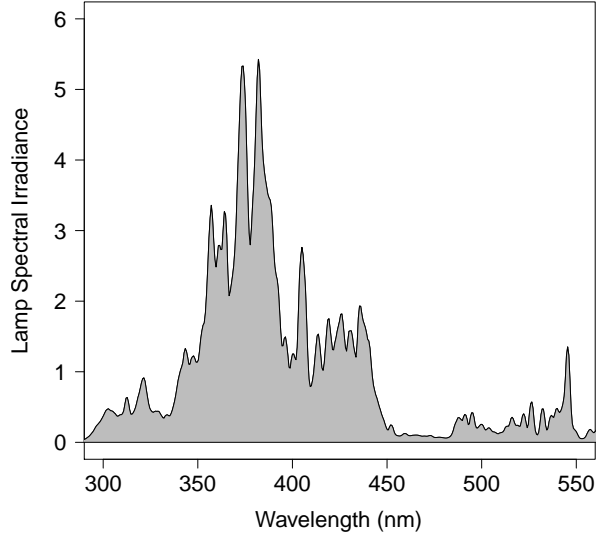}
\end{center}
\caption{Plot of the laboratory accelerated test lamp spectral irradiance distribution.}\label{fig:lamp.spectra}
\end{figure}

The light source for the laboratory accelerated test experiments is
high-intensity UV lamps. The spectral irradiance of the lamps
is a function of wavelength $\lambda$, which gives the power density
at a particular wavelength $\lambda$. The spectral irradiance of the
UV lamps in the NIST SPHERE is illustrated in
Figure~\ref{fig:lamp.spectra}. Specifically, the irradiance is
defined as the power of the electromagnetic radiation per unit area
incident on a surface.

\begin{figure}
\begin{center}
\includegraphics[width=.6\textwidth]{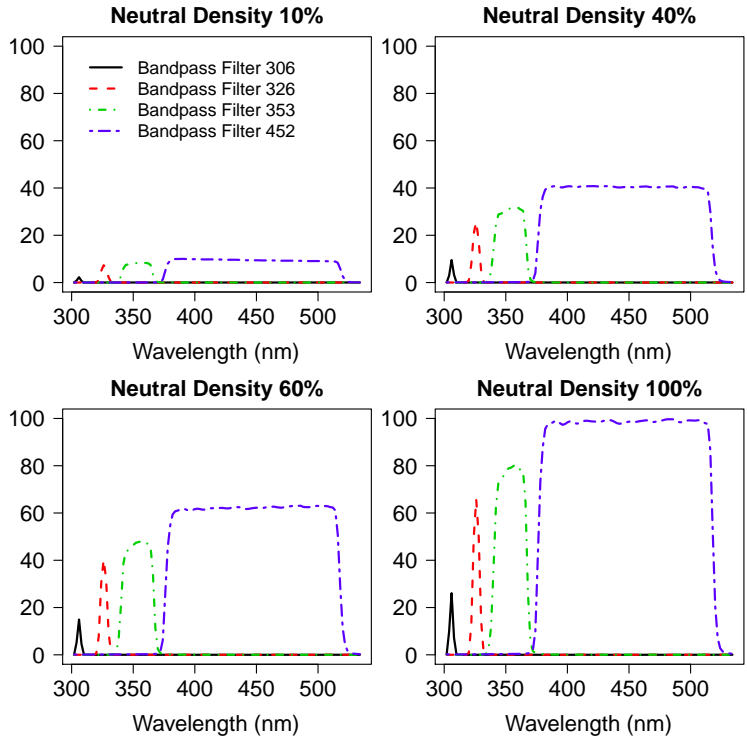}
\end{center}
\caption{Illustration of the combinations of the BP and ND filters. The
  y-axis shows the percentage of photons passing through the combinations of
  filters.}\label{fig:indoor.pb.plot}
\end{figure}

The effect of UV radiation on degradation depends on both the UV
spectrum and UV intensity. UV radiation with shorter wavelengths tend
to have higher energy per photon, thus causing more damage to the
material when compared with UV radiation with longer wavelengths. Also,
for the UV with the same wavelength, higher UV intensity (means more
photons per time unit) tends to cause more damage than lower
intensity. To study the effect of UV spectrum and UV intensity, the
spectral irradiance of the lamps was modified and controlled by
bandpass (BP) and neutral density (ND) filters. BP filters pass only
UV with wavelengths over a particular range. For example, the 306
nanometer (nm) BP filter has a nominal center wavelength of 306~nm
and full-width-half maximum values of $\pm3$~nm. The four BP filters
used in the experiments have nominal center wavelengths of 306~nm,
326~nm, 353~nm, and 452~nm.

ND filters control the intensity of the UV radiation without affecting
the shape of the UV spectrum. For example, a 10\% ND filter
(nominally) passes 10\% of the UV photons at any wavelength. The
four ND filters used in the experiments are 10\%, 40\%, 60\% and
100\% (actually, a 100\% ND would use no ND filter). As an illustration,
Figure~\ref{fig:indoor.pb.plot} shows all combinations of the 16
BP and ND filters.

The laboratory accelerated test experiments also have other
controlled environmental factors: temperature and
RH. Table~\ref{tab:indoor.setup} gives a summary of the experimental
factors for the laboratory accelerated degradation experiment. The
temperature levels were 25$\degreeC$, 35$\degreeC$, 45$\degreeC$,
and 55$\degreeC$. The RH levels were 0\%, 25\%, 50\%, and 75\%. The
laboratory accelerated test data contain a total of 80 combinations
of the experimental factors. Due to time and funding constraints,
not all combinations of the four experimental factors were run in
the experiments. Table~\ref{tab:indoor.tempRH} summarizes the 80
experiment combinations of the BP and ND filters and temperature and RH
levels.  There were four replicates for most of the experimental
factor-level combinations. A total of 319 specimens were exposed in
the laboratory accelerated test experiments.

\begin{table}
\begin{center}
\caption{Laboratory accelerated test setup, showing the BP filters, ND
  filters, and levels of temperature and RH.}\label{tab:indoor.setup}
\vspace{1ex}
\begin{tabular}{c|c}\hline\hline
\multirow{2}{*}{BP filter} & 306~nm ($\pm$3~nm), 326~nm ($\pm$6~nm), \\
&353~nm ($\pm$21~nm), 452~nm ($\pm$79~nm)\\\hline
ND filter & 10\%, 40\%, 60\%, 100\%\\\hline
Temperature  & 25$\degreeC$, 35$\degreeC$, 45$\degreeC$, 55$\degreeC$\\\hline
RH & 0\%, 25\%, 50\%, 75\%\\\hline\hline
\end{tabular}
\end{center}
\end{table}

\begin{table}
\begin{center}
\caption{Summary of the 80 experimental combinations of BP and ND
  filters and temperature and RH levels. An empty cell implies that
  no experiments were done for the corresponding combination of
  temperature and RH. $4\times 4$ implies that experiments were done
  for all of the 16 combinations of the BP and ND filters at the
  corresponding temperature and RH combination. $4\times 1$ implies
  that experiments were done for all four BP filters and the 100\%
  ND filters for the corresponding temperature and RH
  combination.}\label{tab:indoor.tempRH}
\vspace{2ex}
\begin{tabular}{|c|c|c|c|c|}\hline
\backslashbox{Temp}{RH}& $0$\% & $25$\% & $50$\% & $75$\% \\\hline
$25$ & $4\times4$ &            &            &            \\\hline
$35$ & $4\times4$ & $4\times1$ & $4\times1$ &            \\\hline
$45$ &            & $4\times1$ & $4\times1$ & $4\times4$ \\\hline
$55$ &            &            &            & $4\times4$ \\\hline
\end{tabular}
\end{center}
\end{table}

Damage to the material, which is used as an indication for
degradation, was measured by Fourier transform infrared (FTIR)
spectroscopy.  An FTIR spectrometer provides an infrared spectrum of
absorption or emission of a material. In particular, special
structures of compounds absorb the infrared energy at different
wavelengths, which results in peaks in the FTIR spectra. The
locations of the FTIR peaks correspond to unique chemical structures
and thus can be used to identify the relative concentration of
different compounds. The height of a peak is proportional to the
concentration of a particular compound or structure. The time
intervals between the FTIR measurements in the accelerated test were
typically on the order of a few days.

\begin{figure}
\begin{center}
\includegraphics[width=.5\textwidth]{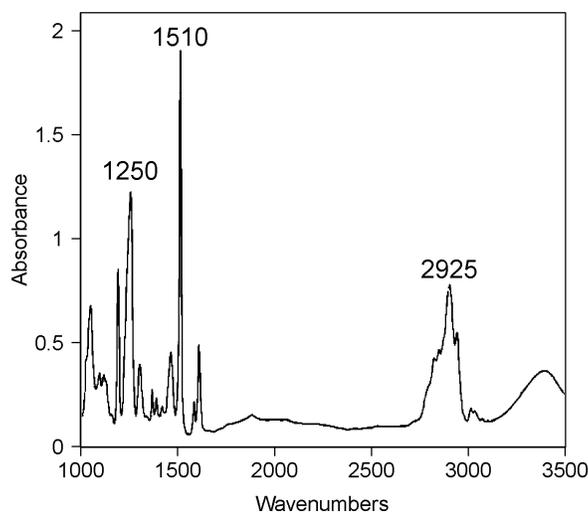}
\end{center}
\caption{Illustration of FTIR spectrum of the model epoxy used in the NIST experiments.}\label{fig:ftir}
\end{figure}

Figure~\ref{fig:ftir} gives an illustration of FTIR peaks for a
particular specimen at one point in time. Our
modeling focuses on intensity changes at wavenumber 1250 $\cm^{-1}$,
which correspond to C-O stretching of aryl ether. Other peaks that
were recorded as potentially useful responses include 1510
$\cm^{-1}$ (benzene ring stretching), 1658 $\cm^{-1}$ (C=O
stretching of oxidation products), and 2925 $\cm^{-1}$ (CH$_2$
stretching) (e.g., see \citeNP{Bellinger1984}, \citeNP{Bellinger1985},
\citeNP{Rabek1995}, and \citeNP{Kelleher1969}).

\begin{figure}
\begin{center}
\includegraphics[width=.55\textwidth]{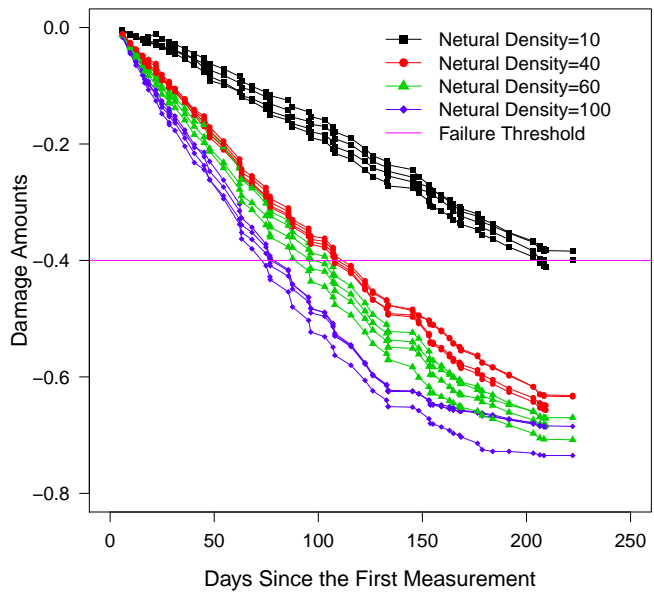}
\end{center}
\caption{Degradation paths for specimens with 10\%, 40\%, 60\% and
  100\% ND filters, the 353~nm  BP filter, temperature at 35$\degreeC$, and 0\% RH.}\label{fig:indoor.sample.paths}
\end{figure}

As an example of the degradation data collected in the laboratory
accelerated test experiments, Figure~\ref{fig:indoor.sample.paths}
shows the degradation paths for FTIR wavenumber 1250 $\cm^{-1}$ for
specimens with 10\%, 40\%, 60\% and 100\% ND filters, the BP filter
centered at 353~nm, temperature 35$\degreeC$, and 0\% RH. For this
wavenumber, the degradation paths are decreasing (i.e., the amount
of C-O stretching of aryl ether was decreasing). As expected, the
degradation rates were higher for the ND filters passing larger
percentages of UV photons. For the groups of two to four specimens
exposed to the same conditions (and at the same time and in the same
chamber), there is some specimen-to-specimen variability.

To use a degradation model to make inferences about failure times,
it is necessary to have a definition of failure. When dealing with
soft failures (as is commonly done in degradation applications),
such definitions generally have a subjective element (e.g., at what
point in loss of gloss of a coating do we have a failure), but such
decisions are typically made in a purposeful manner with great care
(e.g., using customer survey information to assess perception of
gloss loss). These ideas relating degradation modeling to the
estimation of service life are widely used in applications of
degradation data modeling (e.g., the light output of lasers and LEDs,
corrosion of pipelines, and growth of cracks in structures). During
the NIST experimental program, physical measurements of gloss loss
were also taken and correlated with the FTIR chemical degradation
measurements. One reason that we choose to use the wavenumber 1250
$\cm^{-1}$ as our response is that it correlated best with gloss
loss of the model epoxy used in the NIST experiments.  As shown by
the horizontal lines in Figures~\ref{fig:indoor.sample.paths} and
\ref{fig:12.outdoor.sample.path.plot}, a damage level of $-0.40$ was
used as the failure definition.

\subsection{Outdoor-Exposure Experiments and Data}
The UV exposure for the outdoor-exposure specimens is from the
sun. There were 53 specimens in the outdoor-exposure experiments and
they were exposed over different  time intervals during a three-year
period. The UV spectral irradiance, temperature, and RH are, of
course, uncontrolled outside, but were recorded at 12-minute
intervals. For the outdoor-exposure specimens, the UV, temperature, and RH are
dynamic covariates. The measurements of degradation were taken every
three to four days, similar to the accelerated test specimens. We
continue to focus on chemical changes at wavenumber 1250
$\cm^{-1}$. Note, however, that we used the laboratory accelerated test data
for model fitting. The data from the outdoor-exposed specimens are
used only for validating the accelerated test methodology.

\begin{figure}
\begin{center}
\includegraphics[width=.55\textwidth]{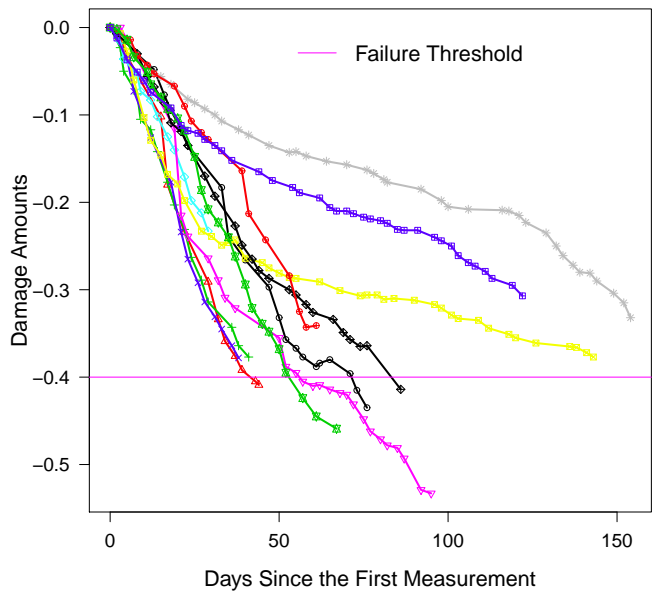}
\end{center}
\caption{Plots of the degradation paths as a function of the days
  since the first measurement for a representative subset of 12 outdoor-exposed specimens.}\label{fig:12.outdoor.sample.path.plot}
\end{figure}

We also want to point out the interesting difference between the
laboratory accelerated test data and outdoor-exposure data. The data
shown in Figure~\ref{fig:indoor.sample.paths} were collected in
laboratory accelerated tests in which the UV, temperature, and RH
are controlled to be constant over time. All of the sample paths
have the same shape.  Figure~\ref{fig:12.outdoor.sample.path.plot},
on the other hand, shows the sample degradation paths as a function
of the days since the beginning of exposure for a representative
subset of 12 specimens that were exposed outdoors at different times. The
sample degradation paths have different shapes, depending on the
time of the year that the specimens were being exposed. The
variability in the shapes of degradation paths for the
outdoor-exposure data is due to variability in the dynamic
covariate time series.

\begin{figure}
\centering
\begin{tabular}{cccc}
\includegraphics[width=.45\textwidth]{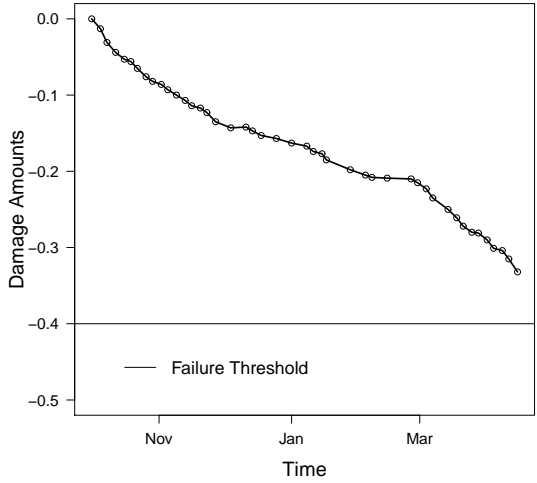}&
\includegraphics[width=.45\textwidth]{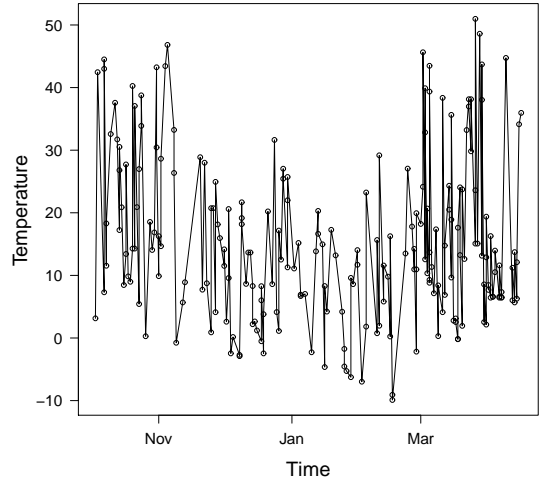}\\
(a) Damage & (b) Temperature\\
\includegraphics[width=.45\textwidth]{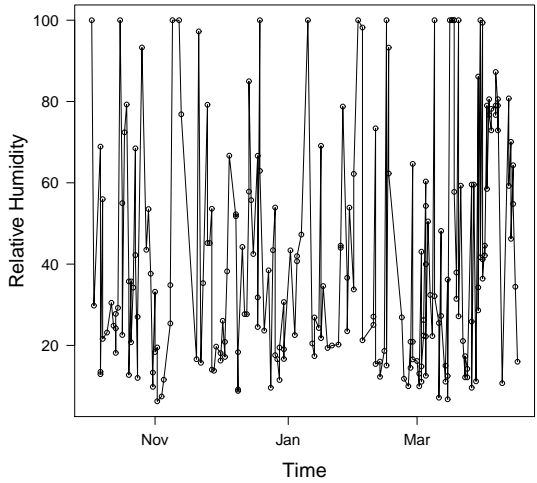}&
\includegraphics[width=.45\textwidth]{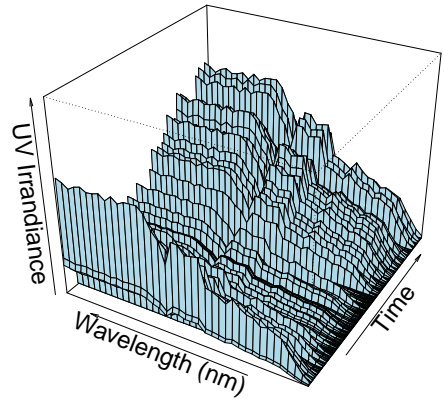}\\
(c) Relative humidity & (d) UV\\
\end{tabular}
\caption{Plots of the degradation path for an outdoor-exposed specimen showing
  the relationship between the degradation and the dynamic
  covariates. (a) the degradation path, b) temperature as a
  function of time, c) RH as a function of time, and d) a
  perspective plot showing the recorded UV intensity as a function of
  time and wavelength.}\label{fig:outdoor.sample.specimen}
\end{figure}

To further illustrate and understand the outdoor-exposure
degradation-path patterns,
Figure~\ref{fig:outdoor.sample.specimen}(a) shows the degradation
path for a particular outdoor-exposed specimen as a function of the
calendar time. Figures~\ref{fig:outdoor.sample.specimen}(b),
\ref{fig:outdoor.sample.specimen}(c), and
\ref{fig:outdoor.sample.specimen}(d) show the dynamic covariates
corresponding to the particular degradation path in
Figure~\ref{fig:outdoor.sample.specimen}(a). From
Figure~\ref{fig:outdoor.sample.specimen}(d), we can see that the UV
intensity is low during the late fall and winter months,
corresponding to a smaller slope in the degradation path, while the
UV is stronger for the months of March and April, corresponding to a
larger slope in the degradation path.

\subsection{Notation}
Here we introduce notation for the data. The degradation (damage)
measurement for specimen~$i$ is the change (relative the value at
the beginning of exposure) in the FTIR peak at 1250 $\cm^{-1}$ at
time $t_{ij}$ and for the laboratory accelerated test data is
denoted by $y_{i}(t_{ij})$, $i=1,\dots,n$, $j=1,\dots,m_i$. Here, $n$
is the total number of laboratory accelerated test specimens and
$m_i$ is the number of time points where the degradation
measurements were taken for specimen~$i$. The last observation time
for specimen~$i$ is denoted by $t_i=t_{im_i}$.

For the laboratory accelerated test data, the UV radiation is
quantified by the cumulative dosage $D_i(\tau_{il})$ at time
$\tau_{il}$.  (Note that the cumulative dosage values were reported
at times that differ from the times at which the degradation
measurements were taken.) The cumulative dosage is proportional to
the total number of photons that were absorbed by specimen~$i$
across all wavelengths between time 0 and $\tau_{il}$. Here,
$i=1,\dots,n$, $l=1,\dots, n_i$, where $n_i$ is the number of time
points at which the total dosage was recorded for specimen~$i$.

For the laboratory accelerated test specimens, the experimental
factors are held constant at specified levels over time. We let
$\BP_i$, $\ND_i$, $\Temp_i$, and $\RH_i$ be the BP filter, ND
filter, temperature, and RH levels, respectively, for
specimen~$i$. In summary, the laboratory accelerated test data are
$\{y_{i}(t_{ij}), D_i(\tau_{il}), \BP_i, \ND_i, \Temp_i, \RH_i\}$
for $i=1,\dots,n$, $j=1,\dots,m_i$, and $l=1,\dots, n_i$.

For the outdoor-exposure data, we use subscript $k$ to index the
exposed specimens. The degradation measurement at time $t_{kj}$ is
denoted by $y_{k}(t_{kj})$, $k=1,\dots,q, j=1,\dots,m_k$ for
specimen~$k$. Here $q$ is the number of outdoor-exposure
specimens. The recorded ambient temperature and RH for specimen~$k$
at time $\tau_{kl}$ are denoted by $\Temp_k(\tau_{kl})$ and
$\RH_{k}(\tau_{kl})$, $l=1,\dots, n_k$, respectively. For the UV
radiation, dosage was recorded for each 12-minute interval and each
2~nm wavelength interval between 300~nm and 532~nm.  We denote the
UV dosage for outdoor-exposure specimen~$k$ at time $\tau_{kl}$ and
wavelength interval $\lambda$ by $D_k(\tau_{kl},\lambda)$.  An
example of $D_k(\tau_{kl},\lambda)$ data is shown in
Figure~\ref{fig:outdoor.sample.specimen}(d). In summary, the
outdoor-exposure data are $\{y_{k}(t_{kj}), D_k(\tau_{kl},\lambda),
\Temp_k(\tau_{kl}), \RH_{k}(\tau_{kl})\}$ for $k=1,\dots,q$,
$l=1,\dots, n_k$, and $j=1,\dots,m_k$.

\subsection{Data Cleaning}
The data required cleaning before the analysis. For a certain number
of specimens in the laboratory accelerated test data, the
degradation paths show two segments instead of continuous
curves. This was believed to have been caused by a specimen-preparation
problem, so those specimens were removed from the dataset. Hence,
we used a total of 302 specimens from the laboratory accelerated test
data for analysis. The degradation paths also show a more
complicated pattern after the damage is below $-0.6$. This behavior was
believed to have been caused by a change in the degradation
mechanism in the specimen. When fitting models using the laboratory
accelerated test data, we use only degradation data above
$-0.6$. Because the failure threshold is $-0.4$ for the
degradation measurement at wavenumber 1250 $\cm^{-1}$, $-0.6$ is far beyond the
definition of failure.

For the outdoor-exposure experiments, temperature and/or RH data for
some time points) were missing. The missing data for temperature and
RH, however, account for a very small percentage of the total
outdoor-exposure data (around 0.32\%). We used observed information
within two weeks of the missing observations to impute replacement
values. For the outdoor-exposure predictions, we used 60-minute
intervals, which is small relative to the total prediction period of
more than 100 days. Thus, there is little sensitivity to the missing
data.

\section{Models for Photodegradation Paths}\label{sec:photo.model}
\subsection{The Concept of UV Dosage}
The UV dosage is an important concept that will be used as the
``time'' scale for the
subsequent photodegradation modeling. For the laboratory accelerated
test data, only the cumulative dosage $D_i(\tau_{il})$ was
available. Conceptually, the cumulative dosage is computed as
follows. The number of incident photons from UV light source, defined
as dose, for specimen~$i$ at time $\tau_{ik}$ from wavelength
$\lambda$ after BP and ND filters, is denoted by
$E_i(\tau_{ik},\lambda)$. Let $\Lamp(\lambda)$ be the spectral
irradiance of the UV lamp as a function of wavelength, and let
$\Filter(\lambda,\BP_i,\ND_i)$ denote the combined effect of the BP
and ND filters. The dose $E_i(\tau_{ik},\lambda)$ can be computed as
\begin{align*}
E_i(\tau_{ik},\lambda)=E_i(\lambda)=\Lamp(\lambda)\times\Filter(\lambda,\BP_i,\ND_i),
\end{align*}
which is constant over time for the laboratory accelerated test specimens
due to the controlled experimental factors. The number of incident
photons absorbed by a specimen at time $\tau_{ik}$, defined as
``dosage,'' is denoted by $D(\tau_{ik},\lambda)$, where
\begin{align*}
D_i(\tau_{ik},\lambda) = E_i(\tau_{ik},\lambda)\{1-\exp[-A(\lambda)]\},
\end{align*}
and $A(\lambda)$ is the spectral absorbance of the specimen at
specified wavelength $\lambda$ (a property of the material).  Thus,
the cumulative dosage, which is proportional to the total number of
photons absorbed by a specimen across all wavelengths up to time
$t$, is computed as $$D_i(t)=\int_{0}^{t}\int_\lambda
D_i(\tau,\lambda)d\lambda d\tau,$$ where the integral is over the
entire range of $\lambda$. We also define
$D_{it}(\lambda)={\int_{0}^t D_i(\tau,\lambda)d\tau}$ to be the
wavelength-specific cumulative dosage.

\subsection{The Physical Model}
To model the effect of the experimental factors, we introduce the
concept of ``effective dosage.'' The cumulative effective dosage up to
time $t$ is defined as
\begin{align}\label{eqn:effective.dosage}
\int_0^t\int_{\lambda_{\min}}^{\lambda_{\max}}D_i(\tau,\lambda)\phi(\lambda)d\lambda d\tau.
\end{align}
Here the function $\phi(\lambda)$ is the quasi-quantum yield
function describing the fact that photons with a shorter
wavelength have a higher probability of causing damage. The
wavelengths that are of interest are between $\lambda_{\min}$ and
$\lambda_{\max}$. For values of $\lambda>\lambda_{\max}$, the
probability of damage is negligible. For values of
$\lambda<\lambda_{\min}$, potentially damaging photons
are normally filtered out by the protective ozone layer in the stratosphere.

To allow for the environmental effects for specimen~$i$, we use
the following model for experimental-variable adjusted effective dosage.
\begin{align}\label{eqn:si(t)}
S_i(t)=\int_0^tf(\Temp_i)g(\RH_i)d(\ND_i)\int_{\lambda_{\min}}^{\lambda_{\max}}D_i(\tau,\lambda)\phi(\lambda)d\lambda d\tau.
\end{align}
Here $f(\Temp_i)$, $g(\RH_i)$ and $d(\ND_i)$ are functions of the
acceleration factors due to temperature, RH, and ND,
respectively.

Dosage $D_i(\tau,\lambda)$ for each specimen was computed taking into
account the nominal values of the ND filters. The percentage of UV
photons passing through the ND filters, however, is not exactly equal
to the nominal values. Thus the factor $d(\ND_i)$ is used to
provide a data-based adjustment for the deviations.

The quasi-quantum yield function $\phi(\lambda)$
describing the effect of UV spectrum is material dependent and
unknown and needs to be estimated from the data. The estimation
of $\phi(\lambda)$ from experimental data helps us understand
material properties and how the UV exposure affects the degradation
process at different wavelengths.

Environmental factors such as temperature and RH will also affect
the degradation process. The Arrhenius relationship is widely used
to describe the rate of chemical reactions and thus the acceleration
effect of temperature. The manner in which RH and UV intensity
(controlled by the neutral density filters) affect the degradation
process, however, is unknown. That is, the functional forms of $g$
and $d$ need to be identified from scientific knowledge of the
degradation process and the experimental data.

\subsection{The Statistical Model for Photodegradation}
In the general degradation path model, the degradation measurement
of specimen~$i$ at time $t_{ij}$ is
\begin{align}\label{eqn:model.yt}
y_{i}(t_{ij})=\Degpath_{i}(t_{ij})+\epsilon_{i}(t_{ij}),
\end{align}
where $\Degpath_{i}(t_{ij})$ is the actual degradation path and
$\epsilon_{i}(t_{ij})$ is the corresponding measurement error.
Photodegradation is primarily driven by the effective dosage
$S_i(t)$ as defined in \eqref{eqn:si(t)}. The general shape of the laboratory accelerated test
degradation paths can be described by the following parametric model
\begin{eqnarray}\label{eqn:overall}
\Degpath_{i}(t_{ij})=\frac{\alpha{\exp(v_{i})}}{1+\exp(-z)}
\end{eqnarray}
where $z=\{\log[S_i(t_{ij})]-\mu\}/\sigma$ and $\mu$ and $\sigma$
are the parameters describing the location and steepness of the
damage curve, respectively. Ignoring the random effect $v_{i}$, the
asymptote $\alpha$ reflects the maximum degradation damage when
total effective dosage goes to infinity. The parameter $\exp(\mu)$
is the half-degradation effective dosage (i.e., the amount of
effective dosage need for the degradation to reach the level
$\alpha/2$). The reciprocal of the scale parameter $1/\sigma$ is
proportional to the slope of the degradation path for any fixed
value of $z$. So a larger value of $1/\sigma$ implies a larger degradation
rate.

In~\eqref{eqn:overall}, the term $v_{i}$ is the individual random
effect for degradation path $i$, which is modeled by a normal
distribution with mean 0. The random effect is used to explain the
specimen-to-specimen variability that is caused by uncontrolled
and/or unobservable factors (e.g., differences in fabricating the
specimens or specimen position in the environmental chamber). The
model in~\eqref{eqn:overall} a nonlinear mixed-effects model. The
statistical literature in this topic is rich. One can refer to, for
example, \citeN{Davidian2003} or \citeN{pinheiro2006mixed} for more
details.

\section{Modeling the Laboratory Accelerated Test Data}\label{sec:modeling.indoor.data}
\subsection{Initial Analysis of Laboratory Accelerated Test Data}
In this section, we perform some initial exploratory analyses of the
laboratory accelerated test data. We start by fitting a
categorical-effects model so that we can study the effects of the
experimental variables without making any apriori assumptions about
the form of the relationships. Because the experimental factors were
held constant over time in the laboratory accelerated test, using the definition of $S_i(t_{ij})$ in \eqref{eqn:si(t)}, the term $z$ in~\eqref{eqn:overall} can be computed as,
\begin{align}\label{eqn:stat.model.const1}
z&=\frac{\log[S_i(t_{ij})]-\mu}{\sigma}\\\nonumber
&=\frac{\log(t_{ij})+\log[b(\BP_i)]+\log[f(\Temp_i)]+\log[g(\RH_i)]+\log[d(\ND_i)]-\mu}{\sigma},
\end{align}
where
\begin{align}\label{eqn:bp.i}
b(\BP_i)=\int_{\lambda_{\min}}^{\lambda_{\max}}\Lamp(\lambda)\Filter(\lambda,\BP_i,\ND_i)\{1-\exp[-A(\lambda)]\}\phi(\lambda){d\lambda}
\end{align}
is the effect of UV spectrum because it integrates over
$\phi(\lambda)$ for the wavelength range defined by the $\BP_i$.

For the model in (\ref{eqn:stat.model.const1}), we use the
constraint that $f(35)=g(25)=d(10)=1$ to ensure that the parameters
are estimable (i.e., we treat temperature 35$\degreeC$, RH 25\%, and
ND 10\% as the baseline experimental setting in the
categorical-effects model). For the UV spectrum effect, only the
values of $\log[b(\BP_i)]-\mu$ are estimable, which is sufficient
because we are only interested in the relative relationship among
the effects of the BP filters. The maximum likelihood (ML) estimates of
parameters in (\ref{eqn:stat.model.const1}) were obtained by using
\texttt{nlme} in R. Degradation paths in a small wavelength interval
       [e.g., $306~\nm$ ($\pm3~\nm$)] have similar steepness. We
       assume $\sigma$ is mainly determined by wavelength. Thus we
       denote the categorical effect by $\sigma_{\lambda}$ for each of the four
       BP filters.

Table~\ref{tab:categorical.effect.par.est} lists the ML estimates of
the fixed-effects parameters in (\ref{eqn:stat.model.const1}).
Although the categorical-effects model only provides estimates of
the UV, temperature, and RH effects at a limited number of points,
the information from the model is useful for
guiding the choice of the functional forms of $\phi(\lambda)$, $d(\ND)$,
$f(\Temp)$, $g(\RH)$, and $\sigma_{\lambda}$ in next modeling stage.

\begin{table}
\begin{center}
\caption{Parameter estimates of the categorical-effects model.}\label{tab:categorical.effect.par.est}
\vspace{1ex}
\begin{tabular}{crrr}\hline\hline
    &    & Standard&\\
  Parameter    &    Estimate& \multicolumn{1}{c}{Error}&$p$-value\\\hline
$\alpha$                 & $-0.6810$  &  0.0130 & $<0.0001$ \\\hline\\[-2.2ex]
$\log[\bar{\phi}(306)]-\mu$    & $-6.5620$  &  0.0755 & $<0.0001$ \\
$\log[\bar{\phi}(326)]-\mu$    & $-7.0844$  &  0.0361 & $<0.0001$ \\
$\log[\bar{\phi}(353)]-\mu$    & $-9.0275$  &  0.0323 & $<0.0001$ \\
$\log[\bar{\phi}(452)]-\mu$    &$-10.1087$  &  0.0350 & $<0.0001$ \\\hline
$\log[d(40)]$            & $-0.7939$  &  0.0201 & $<0.0001$ \\
$\log[d(60)]$            & $-1.0553$  &  0.0199 & $<0.0001$ \\
$\log[d(100)]$           & $-1.3082$  &  0.0200 & $<0.0001$ \\\hline
$\log[f(25)]$            & $-0.1963$  &  0.0092 & $<0.0001$ \\
$\log[f(45)]$            &  0.1973  &  0.0247 & $<0.0001$ \\
$\log[f(55)]$            & $-0.8193$  &  0.0357 & $<0.0001$ \\\hline
$\log[g(0)]$             &  0.8749  &  0.0231 & $<0.0001$ \\
$\log[g(50)]$            & $-0.3707$  &  0.0255 & $<0.0001$ \\
$\log[g(75)]$            &  0.2287  &  0.0240 & $<0.0001$ \\\hline
$\sigma_{306}$            &  1.5591  &  0.0149 & $<0.0001$ \\
$\sigma_{326}$            &  1.2336  &  0.0074 & $<0.0001$ \\
$\sigma_{353}$            &  1.0443  &  0.0057 & $<0.0001$ \\
$\sigma_{452}$            &  0.8416  &  0.0054 & $<0.0001$ \\\hline\hline
\end{tabular}
\end{center}
\end{table}

\begin{figure}
\begin{center}
\includegraphics[width=.5\textwidth]{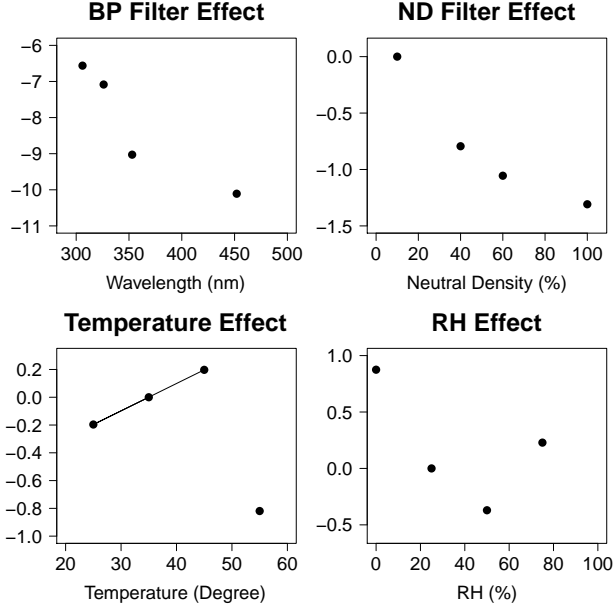}
\end{center}
\caption{ Plots of the categorical effects for UV spectrum, ND filter, temperature, and RH.}\label{fig:categorical.effect}
\end{figure}

\subsection{Effects of the Explanatory Variables}
In this section, we discuss the selection of the functional forms
for the effects of explanatory variables used in the laboratory
accelerated test.

\subsubsection{Modeling the BP Filter Effect}
To suggest a functional form for $\phi(\lambda)$, we initially assume that
$\phi(\lambda)$ is constant over the specific range of each BP
filter, denoted by $\bar{\phi}(\lambda)$. For example, for the 306
$\nm$ BP filter, $\bar{\phi}(306)$ will be used to represent the
effect. From~\eqref{eqn:bp.i}, we obtain
$b(\BP_i)=[D_i(t_i)/t_i]\bar{\phi}(\lambda)$. Because we record
$D_i(t_i)$ and have an estimate of $b(\BP_i)$ from the categorical-effects
model, we can obtain a heuristic estimate for $\bar{\phi}(\lambda)$
from this relationship. For example,
\begin{align}
\bar{\phi}(306)=\frac{b(306)}{(\sum_{i:
    \BP_i=306}[D_i(t_i)/t_i])/(\sum_{i: \BP_i=306} 1)},
\end{align}
for $303~\nm\leq\lambda\leq309~\nm$. Similarly, one can obtain the
estimates of $\bar{\phi}(\lambda)$ for other the BP filters,
$320~\nm\leq\lambda\leq332~\nm$, $332~\nm\leq\lambda\leq374~\nm$,
and $373~\nm\leq\lambda\leq531~\nm$. The corresponding results are
shown in
Table~\ref{tab:categorical.effect.par.est}. Figure~\ref{fig:categorical.effect}(a)
provides a visualization of a simple estimate of
$\phi(\lambda)$. The results suggest that for shorter wavelengths,
there is more damage than at longer wavelengths, agreeing with known
theory. The shape of the curve suggests an exponential relationship.

The quasi-quantum yield $\phi(\lambda)$ describes the fact that
photons at shorter wavelengths have higher energy and thus a higher
probability of causing damage. \citeN{MartinLechnerVarner1994}
say that for polymeric materials, the shape of $\phi(\lambda)$ is typically
exponential decay. The empirical results in our categorical-effects model
also suggest this and thus we use a log-linear function
$\phi(\lambda)=\exp(\beta_0+\beta_{\lambda}\lambda)$, to describe
quasi-quantum yield where $\beta_0$ and $\beta_{\lambda}$ are
parameters to be estimated from the data.

The parameter $\sigma$ in~\eqref{eqn:stat.model.const1} is related
to the slope of the degradation path. Because UV is the main cause
of degradation and shorter wavelength paths tend to have larger
slopes, we model $\sigma$ as a function of $\lambda$. The curve of
categorical-effects estimates of $\sigma_{\lambda}$ versus $\lambda$
suggests an exponential relationship with a lower bound.  Thus we
use the functional form
$\sigma_{\lambda}=\sigma_0+\exp(\sigma_1+\sigma_2\lambda)$ to
describe the effect that UV wavelength has on $\sigma$.

From~\eqref{eqn:stat.model.const1}, one needs to have a wavelength
specific dosage $D_{it}(\lambda)$ to estimate the parameters in
$\phi(\lambda)$ (i.e., $\beta_0$ and $\beta_{\lambda}$). For the
laboratory accelerated test data, however, only the aggregated
dosage $D_i(t)$ data was available. We use an approximate method
to obtain $D_{it}(\lambda)$ from $D_i(t)$. We consider the four
intervals $303~\nm\leq\lambda\leq309~\nm$,
$320~\nm\leq\lambda\leq332~\nm$, $332~\nm\leq\lambda\leq374~\nm$,
and $373~\nm\leq\lambda\leq531~\nm$, corresponding to the four BP
filters. Note that the spectral irradiance after filtering is
$\Lamp(\lambda)\Filter(\lambda,\BP_i,\ND_i)$, and the approximate
trapezoid area under each $\lambda$ interval is denoted as
$\textrm{Area}_{\lambda}$. The integration of
$\textrm{Area}_{\lambda}$ over each of the four wavelength interval is
denoted by $\textrm{Area}_{\bar{\lambda}}$, where $\bar{\lambda}$ is
$306~\nm$, $326~\nm$, $353~\nm$ or $452 \nm$, the BP filter nominal
center points. We define the proportion of area under $\lambda$
relative to its corresponding wavelength range as
$P(\lambda)=\textrm{Area}_{\lambda}/\textrm{Area}_{\bar{\lambda}}$. Note
that
\begin{align}
D_i(t)=\int_{0}^{t}\int_{\lambda}\Lamp(\lambda)\Filter(\lambda,\BP_i,\ND_i)\{1-\exp[-A(\lambda)]\}d\lambda d\tau,
\end{align}
and the specific form of $A(\lambda)$ is unknown. For the narrow
intervals $303~\nm\leq\lambda\leq309~\nm$ and
$320~\nm\leq\lambda\leq332~\nm$, we can assume
$\{1-\exp[-A(\lambda)]\}$ is constant because the fluctuation over
the narrow range of wavelengths is relatively small. Thus, for
$303~\nm\leq\lambda\leq309~\nm$ and $320~\nm\leq\lambda\leq332~\nm$,
we obtain approximate values from
$D_{it}(\lambda)=D_i(t)P(\lambda).$ Although the interval
$373~\nm\leq \lambda \leq 531~\nm$ is wide, the variation in
$D_{it}(\lambda)$ will be small because both
$\{1-\exp[-A(\lambda)]\}$ and $\phi(\lambda)$ are small over the
interval. Thus we can assume that $D_{it}(\lambda)$ is constant for
$373~\nm\leq \lambda \leq 531~\nm$. For the
$332~\nm\leq\lambda\leq374~\nm$, the lamp spectra curve is
complicated and $\{1-\exp[-A(\lambda)]\}$ is typically not small
enough to do a trapezoid approximation. Thus, in the subsequent
modeling, we (as in the categorical-effects model) use
$\bar{\phi}(353)$ to represent the effect of the 353 $\nm$ BP filter
and treat it as an unknown parameter to be estimated from the
data. There is, however, enough information from the other three BP
filters for us to estimate the unknown parameters of the log-linear
relationship for $\phi(\lambda)$.

\subsubsection{ND Filter Effect}
A power law relationship is typically used to describe the ND effect
(e.g., \citeNP{James1977}). The power law relationship is based on
Schwarzschild's law, which says that the photo-response of radiation
over a given time period has a form $\ND^p$, where $\ND$ is the UV
intensity level. To achieve the same photo-response, $\ND^p\times t$
should be the same, where $p$ is the Schwarzschild coefficient and
$t$ is the exposure time. When $p=1$, this relationship is called
the reciprocity law. Experimental deviations from the reciprocity
law are called reciprocity law failure. More discussion about
Schwarzschild's law and reciprocity can be found in
\citeN{MartinChinNguyen2003}.

Figure~\ref{fig:categorical.effect}(b)
shows the effects of the ND filter. A power law relationship
$d(\ND_i)={\ND_i}^p$ gives a perfect fit to the four points.  Note
that the $\Filter(\lambda,\BP_i,\ND_i)$ already includes the effect
of the ND filter as ${\ND_i}$ with a power of one. Thus, the overall
effect of ND filter is ${\ND_i}^{(1+p)}$. If the reciprocity law
(i.e., the effect of ND is ${\ND_i}^1$) holds, $p$ should be equal
to zero in this parameterization. Thus, combining the physical
knowledge and the empirical evidence, we used the power law
relationship to describe the UV intensity effects. Another way of
thinking about this is that with the reparameterization, the effect
$p$ describes the deviation between nominal properties of the ND
filters and the actual amount of photon attenuation provided by the ND
filters.

\subsubsection{Temperature Effect}
Figure~\ref{fig:categorical.effect}(c) shows the effect that
temperature has on the degradation rate.  The Arrhenius relationship
is widely used to describe the acceleration effect of temperature on
the rate of a chemical reaction (e.g.,
\citeNP{MeekerEscobarLu1998}). According to the Arrhenius
relationship, the logarithm of the reaction rate should be
proportional to reciprocal temperature in the Kelvin
scale. In particular, the Arrhenius
relationship is
\begin{align}\label{eqn:arrhenius}
f(\Temp_i)&=\gamma_{0}\exp\left(\frac{-E_a/R}{\textrm{TempK}_i}\right),
\end{align}
where $\textrm{TempK}_i$ is the Kelvin temperature computed as
Celsius temperature plus 273.15, $E_a$ is the effective activation
energy, and $R$ is a gas constant. We define $E_a/R$ to be the
temperature effect to be estimated from the data.  The
categorical-effects estimates agree well with this relationship except for the
specimens at 55$\degreeC$ and 75 \%RH.

A possible explanation for the change in the estimated temperature effect at
55$\degreeC$ is that there is an interaction between the high temperature and the
high RH level. Such an interaction could arise because water
release is known to affect the rate of degradation. We can not,
however, estimate the interaction effect because there is data at
55$\degreeC$ for only one RH level. Another possible explanation is
that there had been a failure of an integrated circuit chip in a
controller that caused certain chambers to be overheated for a
period of time. This could have lead to a different
failure mechanism for the affected specimens. Based on these
considerations, we still use the Arrhenius relationship to model the
temperature effect after removing the data at 55$\degreeC$ and 75\% RH.

\subsubsection{RH Effect}
The effect of relative humidity on coating degradation is
complicated. There are few theoretical results to suggest the
functional form for humidity effect in this type of application. It
is known that low humidity will accelerate the side-chain scission
process. As more end groups are created, the degradation rate will tend
to increase. On the other hand, higher water content in the coating
(caused by higher levels of RH) will tend to increase the diffusion
rate of oxygen in the oxidation zone, which can also increase the
degradation rate (e.g., \citeNP{ChenFuller2009}, and
\citeNP{Kiil2012}). Thus, there is a middle range of RH values where
the degradation rate would be expected to be smaller than at the
extremes. These mechanisms suggest a hump shape function for the
effect that RH has on degradation.
Figure~\ref{fig:categorical.effect}(d) shows the categorical-effects
model estimates
for the  RH effect. The effect is increasing first and then decreasing,
suggesting a concave relationship. Based on the empirical evidence
and the suspected chemical reaction mechanisms, we used a quadratic model
\begin{align}\label{eqn:rheffect}
\log[g(\RH)]&=-\beta_{\RH}(\RH-\rh_0)^2
\end{align}
 to describe the RH effect. Here, $\beta_{\RH}$ and $\rh_0$ are
 unknown parameters to be estimated from the data.

\subsection{The Combined Model}
Combining all of the identified functional forms for the effects of the
experimental variables gives following model for the underlying degradation path,
\begin{align}\label{eqn:funcfor}
\Degpath_{i}(t_{ij})&=\frac{\alpha{\exp({v_{i}})}}{1+\exp(-z)},
\end{align}
where
\begin{align*}
z&=\frac{\eta_0+\log[D_i(t_{ij})]+A+p(\log[\ND_i])-\left(\frac{E_a/R}{\textrm{TempK}_i}\right)-
\beta_{\RH}\left(\RH_i-\rh_0\right)^2}{\sigma_0+\exp(\sigma_1+\sigma_2\lambda)},\\
A &= \log\left[\int_{\lambda_{\min}}^{\lambda_{\max}}P(\lambda)\exp(\beta_{\lambda}\lambda) d\lambda\right],
\end{align*}
and $v_i$ is the random effect. Note that the total effective dosage
for wavelength $\lambda$ is $S_i(t,
\lambda)=D_{it}(\lambda)\exp(\beta_0+\beta_{\lambda}\lambda)$, which
is proportional to $D_{it}(\lambda)\exp(\beta_{\lambda}\lambda)$. We
use $D_i(t)\times P(\lambda)\times\exp(\beta_{\lambda}\lambda)$ to
approximate $D_{it}(\lambda)\exp(\beta_{\lambda}\lambda)$. We define
the constant $\eta_0=\beta_0+\log(\gamma_0)-\mu$ because the
$\mu$ and the individual intercept terms are not
independently estimable in the model.

\begin{figure}
\begin{center}
\includegraphics[width=.55\textwidth]{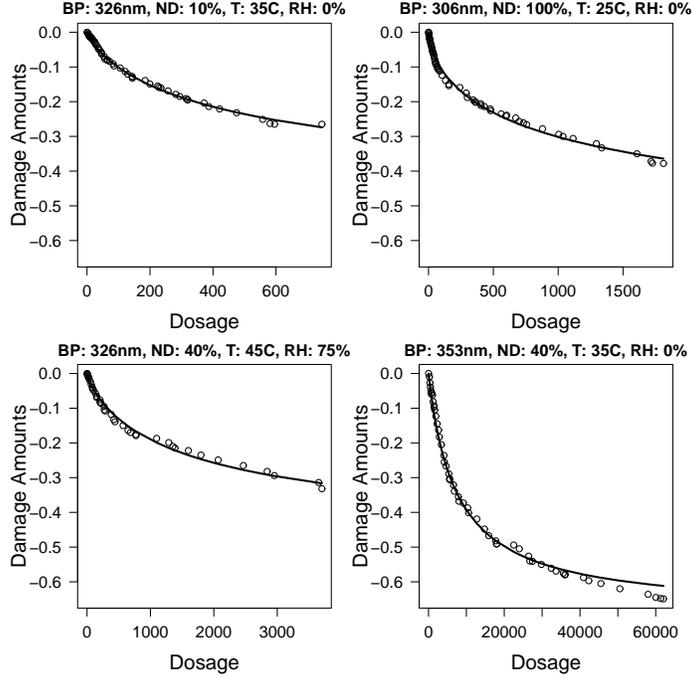}
\end{center}
\caption{Fitted degradation paths for four randomly selected
  specimens based on the model in~\eqref{eqn:funcfor}. The points
  are the measured values and the lines show the fitted values. The
  plot titles show the levels of the experimental
  factors.}\label{fig:fittedindoor}
\end{figure}

Table~\ref{tab:functionalform.effect.par.est} lists the ML estimates
of the parameters in (\ref{eqn:funcfor}). The maximum degradation
damage when total effective dosage goes to infinity is $-0.6191$,
not considering random effects. For the ND filter effect, the power $p$
is estimated to be $-0.5606$, which is significantly different from
0. Thus there is evidence that the reciprocity law does not hold in
this application. The combined ND effect in
$\Filter(\lambda,\BP_i,\ND_i)$ is $1-0.5606=0.4394$. That is, $\ND^{0.4394}$
describes the overall effect of the ND filters. For example, the effect
of a nominal 80\% $\ND$ filter is $100(0.80^{0.4394})$\% = 90.6\%
filtering. As expected, the quasi-quantum yield coefficient
$\beta_{\lambda}=-0.0297 < 0$ indicating that shorter wavelengths
cause more damage. Figure~\ref{fig:fittedindoor} shows examples of
our model~(\ref{eqn:funcfor}) fitted to the laboratory accelerated test
data, showing good agreement.

\begin{table}
\begin{center}
\caption{Parameter estimates for the combined model in~\eqref{eqn:funcfor}.}\label{tab:functionalform.effect.par.est}
\vspace{1ex}
\begin{tabular}{crrr}\hline\hline
      &    & Standard&\\
 Parameter     &  Estimate& \multicolumn{1}{c}{Error}&$p$-value\\\hline
$\alpha$              &  $-0.6191$ & 0.01013 & $<0.0001$ \\\hline
$\beta_{\lambda}$     &  $-0.0297$ & 0.00026 & $<0.0001$ \\
$p$                   &  $-0.5606$ & 0.00781 & $<0.0001$ \\
$\frac{E_a}{R}$       &1945.6482 &75.83458 & $<0.0001$ \\
$\beta_{\RH}$         &  $-0.0005$ & 0.00001 & $<0.0001$ \\
$\rh_0$               &  45.4748 & 0.28749 & $<0.0001$ \\\hline
$\eta_0$              &   9.8986 & 0.25662 & $<0.0001$ \\
$b(353)$              & $-11.5661$ & 0.09428 & $<0.0001$ \\
$\sigma_0$            &   0.8019 & 0.00664 & $<0.0001$ \\
$\sigma_1$            &   7.6776 & 0.18760 & $<0.0001$ \\
$\sigma_2$            &  $-0.0260$ & 0.00062 & $<0.0001$ \\\hline\hline
\end{tabular}
\end{center}
\end{table}

\section{The Prediction Model for Outdoor-Exposure Data}\label{sec:pred.outdoor}
In this section, we adapt the laboratory accelerated test
model~(\ref{eqn:funcfor}) and its parameter estimates to predict
outdoor-exposure degradation.
\subsection{The Cumulative Damage Model for Outdoor-Exposure Degradation Prediction}
For computational convenience, we used 60 minutes instead of 12
minutes as the time interval for the dynamic covariates. For
outdoor-exposure specimen~$k$, we define the incremental effective
dosage at wavelength interval $\lambda \pm 1$ over the 60-minute
interval starting at $\tau$ to be
\begin{align}
\Delta S_k^{\ast}(\tau,{\lambda})&=\int_{\tau}^{\tau+60\,\min}\int_{{\lambda}-1\,\nm}^{{\lambda}+1\,\nm}D_k(\tau,\lambda)\exp(\beta_{\lambda}\lambda)d\lambda d\tau.
\end{align}
Here we use an ``$\ast$'' to indicate that the difference from the
effective dosage defined previously. The previous definition used
$\phi(\lambda)$ but here we use $\exp(\beta_{\lambda}\lambda)$,
which is proportional to $\phi(\lambda)$. The effective dosage
across all wavelengths at time $\tau$ is
$S_{k\tau}^{\ast}(\tau)=\int_{\lambda}\Delta
S_k^{\ast}(\tau,{\lambda})d\lambda$. The cumulative total effective
dosage across all wavelengths from time 0 to time $t$ is
$S_k^{\ast}(t)=\int_{0}^{t}S_{k\tau}^{\ast}(\tau)d\tau$. Temperature
and RH are averaged over all 60-minute intervals. Because no ND
filters were used during the outdoor exposures, we set ND to be
100\% for all outdoor-exposure predictions.

According to the cumulative damage model, the slope of the degradation curve at time $\tau$ and wavelength $\lambda$ is a function of total effective dosage $S_k^{\ast}(\tau)$ and other environmental effects. That is,
\begin{eqnarray}\label{eqn:cumuslope}
g_k'(\tau,\lambda)=\frac{d\Degpath_k(\tau)}{d[S_k^{\ast}(\tau)]}=\frac{1}{S_k^{\ast}(\tau)\sigma_{\lambda}}\times\frac{\alpha\exp(z)}{[1+\exp(z)]^2},
\end{eqnarray}
where
\begin{eqnarray}\label{eqn:finaldefz}
z=\frac{\log[S_k^{\ast}(\tau)]+\eta_0+p[\log(\ND)]-\left[\frac{E_a/R}{\Temp_k(\tau)+273.15 }\right]-\beta_{\RH}\left[\RH_k(\tau)-\rh_0\right]^2}{\sigma_0+\exp\left(\sigma_1+\sigma_2\lambda\right)}\,.
\end{eqnarray}
Note that here we compute the slope $g_k'(\tau,\lambda)$ as a
function of $\tau$ and $\lambda$ because $\sigma_{\lambda}$ depends
on $\lambda$ and the incremental damage amounts need to be
accumulated across the time $\tau$ and wavelength $\lambda$
intervals. The incremental damage, $\Delta\Degpath_k(\tau,\lambda)$,
is the damage at time $\tau$ that was caused by the UV radiation in
the $2 \nm$ wavelength interval $(\lambda-1,\lambda+1)$. In
particular, $\Delta\Degpath_k(\tau,\lambda)=g_k'(\tau,\lambda)\Delta
S_k^{\ast}(\tau,{\lambda})$. The additivity law is assumed, implying
that the damage can be summed up from each wavelength interval in
every 60-minute time interval. Then $\Delta \Degpath_k(\tau)$
denotes incremental damage at time $\tau$ from all wavelengths,
$\Delta\Degpath_k(\tau)=\sum_{\lambda}\Delta\Degpath_k(\tau,\lambda)$. The
cumulative damage $\Degpath_k(t)$ from time 0 to $t$ from all
wavelengths is
\begin{align}\label{eqn:outdoor.pred}
\Degpath_k(t)=\sum_{\tau=0}^{t}\Delta\Degpath_k(\tau).
\end{align}
Hence, degradation $\Degpath_k(t)$ can be predicted  based on the model
estimated from the laboratory accelerated test data. Because there
is a random effect $v_k$ in the mean structure $\Degpath_k(t)$, for the
point prediction, we set $v_k$ to be zero when computing
point predictions.

\subsection{Outdoor-Exposure Prediction Uncertainty Quantification}
The outdoor-exposure prediction involves two sources of variability: the
random effect $v_k$ and the variability in $\thetavechat$. Here we
use $\thetavec$ to denote all the parameters in~\eqref{eqn:funcfor}, and $\thetavechat$ is the ML estimator. The
corresponding variance-covariance matrix is denoted by
$\Sigma_{\thetavechat}$. We use prediction intervals to quantify the
prediction uncertainty. Prediction intervals are calculated and
calibrated following a procedure that is similar to those described
in \citeN{HongMeekerMcCalley2009}, using the
\citeN{LawlessFredette2005} predictive distribution. For notational
simplicity, let $\Degpath=\Degpath_k(t)$, because we compute
pointwise prediction intervals. The cumulative distribution function
of $\Degpath$ at a particular point in time $t$ is denoted by
$F(\Degpath;\thetavec)$ which is primarily determined by the
distribution of the random effect. In particular, the algorithm to
compute the predictive distribution is,

\begin{inparaenum}
\item Simulate $B$ sample estimates $\thetavechat^{\ast}_{b} \sim \NOR(\thetavechat, \Sigma_{\thetavechat})$ and $v_{b}^{\ast}\sim \NOR(0,\widehat{\sigma}^{2}_{v}),  b=1,\dots,B$. We use $B=50{,}000$.

\item Compute the degradation $\Degpath^{\ast}_{b}, b=1,\dots,B$
  using the method summarized by~\eqref{eqn:outdoor.pred} under parameter
  $\thetavechat^{\ast}_{b}$ and the random effect $v_{b}^{\ast}$.

\item Compute $W^{\ast}_{b}=F(\Degpath^{\ast}_{b}|\thetavechat_{b}^{\ast}), b=1,\dots,B$.

\item Compute $w^{l}$ and $w^{u},$ the lower and upper $\alpha/2$ sample
  quantiles, respectively, of $W^{\ast}_{b}$.

\item
Solve $F(\Degpath^{l}|\thetavechat)=w^{l},
F(\Degpath^{u}|\thetavechat)=w^{u}$ for $(\Degpath^{l},
\Degpath^{u})$, providing the $100(1-\alpha)\%$ calibrated
prediction interval.
\end{inparaenum}

This algorithm needs to be repeated over the range of $t$ values of interest.

\subsection{Outdoor-Exposure Prediction Results and Model Comparisons}
\subsubsection{Outdoor-Exposure Predictions}
\begin{figure}
\begin{center}
\includegraphics[width=.6\textwidth]{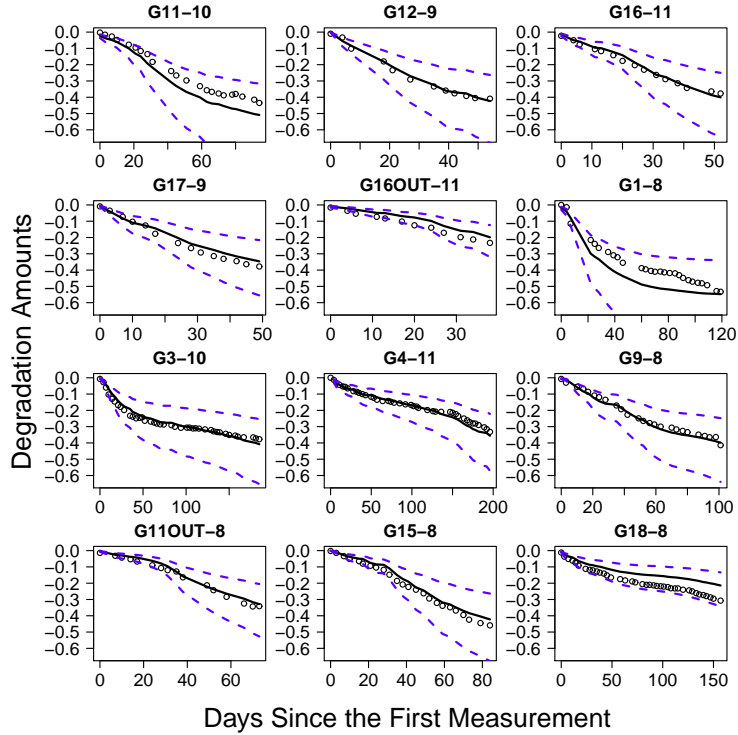}
\end{center}
\caption{Prediction results for 12 representative outdoor-exposed
  specimens. The points show the measured values, the
  solid lines show the predicted values, and the dashed lines show
  the 95\% pointwise prediction intervals.}\label{fig:fittedoutdoor}
\end{figure}

\begin{figure}
\begin{center}
\includegraphics[width=0.5\textwidth]{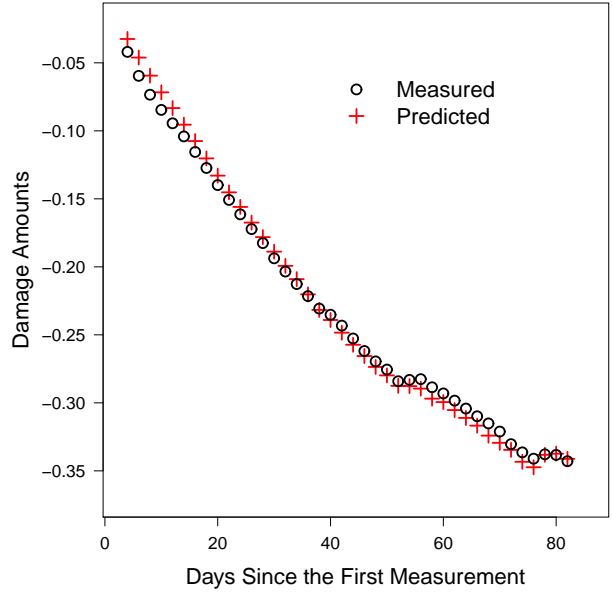}
\end{center}
\caption{Plots of averaged outdoor-exposed degradation measurements and
  values predicted by the cumulative damage
  model.}\label{fig:fittedoutdooraverage}
\end{figure}

Figure~\ref{fig:fittedoutdoor} compares the measured and predicted
degradation paths based on our cumulative damage model for the same
representative set of outdoor-exposed sample paths shown in
Figure~\ref{fig:12.outdoor.sample.path.plot}. The predicted values for
some specimens agree well with the measured values, while for others
the predicted values are either above or below that of the actual
outdoor-exposed sample paths. These variations correspond to the
distribution of the random effects. Most of the measured data points
are within the calibrated prediction intervals, except for some
small levels of degradation at early times which may have been caused
by measurement error. Because the random effects are modeled as
normally distributed with mean $0$, the average predicted values
should be close to the averaged measured values for all 53
outdoor-exposed specimens. Figure~\ref{fig:fittedoutdooraverage} shows the
average of predicted and measured damage for all of the
outdoor-exposed specimens. The average predicted values correspond
well to average measured values.

We saw that the random effects tend to be similar within the same
outdoor-exposed group. For example, four specimens from
outdoor-exposed group G1 all have predictions larger than the
measured values. The four specimens from group G16OUT all have
predictions smaller than the measured values, and the four specimens
from group G4 all have predictions close to the measured
values. These suggest that the random effects could be related to
group conditions such as additional weather-related effects not accounted
for in our model. Other factors
that may contribute to the random effects include the non-uniform
spatial irradiance of specimens, possible non-uniformity of the
material of the specimens, etc.

\subsubsection{Predictions with Early Degradation Information}
\begin{figure}
\begin{center}
\includegraphics[width=.6\textwidth]{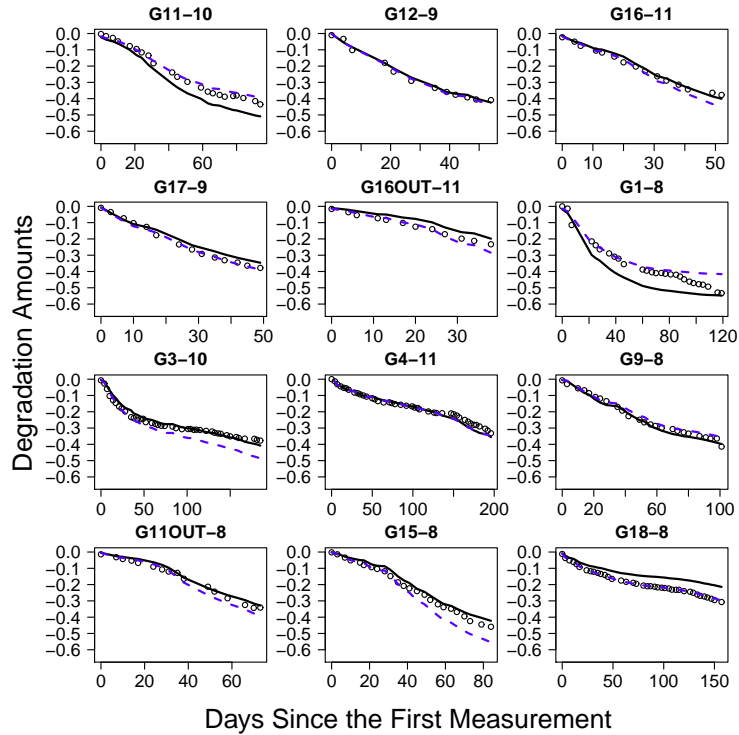}
\end{center}
\caption{Prediction results for 12 representative outdoor-exposed specimens,
  with adjustment made by random effects estimated from the 5th to
  10th data points. The points show the measured values, the
  solid lines show the predictions without adjustments, and the
  dashed lines show the predictions with
  adjustments.}\label{fig:estrandeffect}
\end{figure}

When information about the early part of degradation path is
available for a particular specimen~$k$, that information can be used to
estimate the random effect $v_k$, providing a more precise prediction
of the future amount of degradation. Such predictions are often
needed in practice (say for a fleet of units in the field or for
individual units) to estimate the distribution of remaining life.
In particular, we use the fifth to tenth data points and use the
least squares approach to find $\widehat{v}_k$. That is
$\widehat{v}_k$ is the value that minimizes
$\sum_{j=5}^{10}[y_{kj}-\exp(v_k)\widehat{y}_{kj}]^2$, where
$\widehat{y}_{kj}$ is obtained by substituting the ML estimates
into the prediction model. Then the predicted path for specimen~$k$
is obtained as $\exp(\widehat{v}_k)\widehat{y}_{kj}$. The first four
data points were not used because their damage values were
too small (i.e., the damage amount is less than 0.01) to be useful
in estimating $v_k$. Figure~\ref{fig:estrandeffect} shows the
results for several example specimens where we estimated random
effect $\exp(v_k)$ using the early part of the degradation. The
dashed lines indicate the predicted values after adjusting. For most
specimens, the adjusted predicted values match the measured values
considerably better than the unadjusted values.

\subsubsection{Comparisons}
This section describes comparisons among several models. We use
the Akaike information criterion (AIC) for model-fitting comparisons and
mean squared error (MSE) for prediction comparisons. We
considered the following models for comparisons.

\begin{itemize}
\item Model A: A model similar to that used in
  \citeN{VacaTrigoMeeker2009}, using no
  random effect and where UV intensity was not
  modeled directly.

\item Model B: The model in \eqref{eqn:overall} with individual
  random effects for each specimen and carefully modeled effects for all of the
  experimental variables. For
  predictions from this model, there are two variants.
    \begin{itemize}
    \item Model B1: Prediction with all random effects set equal to
      the expected value of zero.
    \item Model B2: Prediction using some of the early data points
      to estimate the random effects for the individual specimens.
    \end{itemize}
\item Model C: The model in \eqref{eqn:overall} can be easily extended to more complicated random-effects structures. For Model C, we consider the model in \eqref{eqn:overall}  but with both specimen-to-specimen
  and group-to-group random effects. Note that there were typically
  four replicates within experimental group (i.e., exposed at the
  same time and in the same chamber).  For predictions from this model, we also
  have two variants.
    \begin{itemize}
    \item Model C1: Prediction with all random effects set equal to
      the expected value of zero.
    \item Model C2: Prediction using some of the early data points
      to estimate the random effects for the groups and the individual specimens.
    \end{itemize}
\end{itemize}
Table~\ref{tab:model.comparisons} shows the model comparison
results. The results show that the proposed Models~B and~C provide a
much better fit to the laboratory accelerated test data than the
model in \citeN{VacaTrigoMeeker2009}. There is not much difference
between the prediction performance of Models~B and~C. Both models provide
much better predictions than Model~A.

\begin{table}
\begin{center}
\caption{Comparisons of model fits and
  predictions.}\label{tab:model.comparisons}
\vspace{1ex}
\begin{tabular}{c|c|c|c|c|c}\hline\hline
\multirow{2}{*}{Model}	& Log likelihood & Number of & \multirow{2}{*}{AIC} & \multicolumn{2}{c}{Prediction} \\\cline{5-6}
     	& values & parameters &  & Model & MSE \\\hline
A	    & 15740.22             &  9          &	$-31462.44$  & A &      0.004238  \\\hline
\multirow{2}{*}{B}	    & \multirow{ 2}{*}{30201.98}  & \multirow{ 2}{*}{13}          &	\multirow{ 2}{*}{$-60377.95$}	 & B1&	    0.002879  \\\cline{5-6}
        &                      &             &               & B2&   	0.002522  \\\hline
\multirow{2}{*}{C}	    & \multirow{ 2}{*}{30414.43}	         & \multirow{ 2}{*}{14}          &	\multirow{ 2}{*}{$-60800.85$}	 & C1&  	0.002524  \\\cline{5-6}
        &                      &	         &		         & C2&  	0.002396  \\\hline\hline
\end{tabular}
\end{center}
\end{table}

\section{Conclusions and Areas for Future Research}\label{sec:conclusion2}
This paper describes the development of an accelerated test
methodology for photodegradation, including a predictive model that
uses laboratory accelerated degradation test data to predict the service
life of specimens subjected to outdoor exposure. The methodology was
verified by predicting damage for similar specimens that were
exposed outdoors. We developed a physically motivated nonlinear
regression model with random effects to describe the laboratory
accelerated test degradation data, carefully studied the functional
forms of the experimental variables to develop the model, and
estimated model parameters from the accelerated test data. Then we used
a cumulative damage model, incorporating the parameter estimates
from the laboratory accelerated test and individual specimen dynamic
covariate information, to predict the individual outdoor-exposed
degradation paths. We also developed an algorithm to calculate the
prediction intervals. In addition, we showed how to estimate the
specimen-to-specimen random effect for an individual specimen,
providing a means of predicting remaining service life for a unit
that has been in service.

The degradation modeling and prediction methods presented in this
paper serve as an important step in the development of the science
of outdoor weathering service life prediction. There are, however,
several areas for further research.

\begin{itemize}
\item
Given a probability model for degradation paths (such as the one
developed in this paper) and corresponding random effects, a
specific set of dynamic-covariate time series, and a definition of
the corresponding soft-failure threshold, it is possible to compute
the failure-time distribution for exposed units. Chapter 13 of
\citeN{MeekerEscobar1998} illustrates this for simple
constant-environment situations.  The ideas there could be
generalized to compute an estimate of a failure-time distribution
for a specified dynamic-covariate history. In general, evaluation
will require simulation.
\item
A further extension could use a model for the dynamic covariates
(similar to that used in \shortciteNP{HongDuanMeekerGuStanley2015}) to find a
failure-time distribution that takes into account uncertainty in the
future realizations of the dynamic covariates. Models with autocorrelation for the error terms in \eqref{eqn:model.yt} can also be considered.

\item
The predictions generated in this paper and the corresponding
failure-time distributions mentioned above correspond to a given
``weather'' realization from the model for the weather which might
be a ``typical'' realization or, perhaps, a harsher one, to get more
conservative predictions. If there were a need to predict actual
failures for a population of product units in the field, then one
would have to consider a mixture of different weather models and
generate predictions for each, weighted by the amount of product
subject to each such weather model.

\item
In general,
outdoor environments are complicated. In the NIST outdoor-exposure
experiments, the specimens were placed in a covered chamber so that the
main driving factors were UV, temperature, and RH. A more extensive
experiment could be conducted to study the effect of factors like
contaminants in the air, dust, acid rain, and extreme events.

\end{itemize}

\section*{Acknowledgments}
The authors thank the editor, an associate editor, and the referees, for their valuable comments that lead to significant improvement on this paper. The authors acknowledge Advanced Research Computing at Virginia Tech for providing computational resources. The work by Hong was partially supported by the National Science Foundation under Grant CMMI-1634867 to Virginia Tech.

\bibliographystyle{chicago}
\bibliography{ref}
\end{document}